\documentclass[a4paper,12pt]{article}
\pdfoutput=1
\usepackage[intlimits,centertags]{amsmath}
\usepackage{amssymb,amsfonts}
\usepackage{bbm}
\usepackage{mathrsfs}
\usepackage{cite}
\usepackage[scale={0.78,0.8}]{geometry}
\usepackage{multirow}
\usepackage{relsize}
\usepackage[subrefformat=parens,labelformat=parens]{subfig}
\usepackage{xcolor}
\usepackage{caption}
\usepackage{cancel}
\usepackage{graphics,graphicx}
\usepackage{float}
\numberwithin{equation}{section}

\def\beq{\begin{align}}
\def\eeq{\end{align}}
\def\beqa{\begin{eqnarray}}
\def\eeqa{\end{eqnarray}}

\begin{document}

\pagestyle{empty}
\rightline{IFT-UAM/CSIC-17-014}
\vspace{1.2cm}

\vskip 1.5cm

\begin{center}
\LARGE{$R+\alpha R^n$ Inflation in higher-dimensional Space-times}
\\[13mm]
  \large{Santiago Paj\'{o}n Otero$^{1}$, Francisco G. Pedro$^{1}$, and Clemens Wieck$^{1}$  \\[6mm]}
\small{
${}^1$Departamento de F\'{\i}sica Te\'orica
and Instituto de F\'{\i}sica Te\'orica UAM/CSIC,\\[-0.3em]
Universidad Aut\'onoma de Madrid,
Cantoblanco, 28049 Madrid, Spain
\\[8mm]}
\end{center}
\vspace{2cm}
\begin{abstract}
We generalise Starobinsky's model of inflation to space-times with $D>4$ dimensions, where $D-4$ dimensions are compactified on a suitable manifold. The $D$-dimensional action features Einstein-Hilbert gravity, a higher-order curvature term, a cosmological constant, and potential contributions from fluxes in the compact dimensions. 
The existence of a stable flat direction in the four-dimensional EFT implies that the power of space-time curvature, $n$, and the rank of the compact space fluxes, $p$, are constrained via $n=p=D/2$. Whenever these constraints are satisfied, a consistent single-field inflation model can be built into this setup, where the inflaton field is the same as in the four-dimensional Starobinsky model. The resulting predictions for the CMB observables are nearly indistinguishable from those of the latter.
\end{abstract}
\newpage

\setcounter{page}{1}
\pagestyle{plain}
\renewcommand{\thefootnote}{\arabic{footnote}}
\setcounter{footnote}{0}

\tableofcontents

%

\section{Introduction}

As one of the first examples of single-field slow-roll inflation, Starobinsky proposed a model of extended gravity with $f(R) = R + \alpha R^2$ that leads to a scalar field theory with an exponentially flat potential \cite{Starobinsky:1980te}. By means of a Legendre-Weyl transformation the non-trivial gravity action can be recast into the form of Einstein-Hilbert gravity with a minimally coupled scalar field, $\phi$, whose scalar potential takes the form
\begin{align}\label{starobinsky}
V = \frac{1}{8\alpha}\left(1-e^{-\sqrt{2/3} \phi}\right)^2\,.
\end{align}
This model of inflation is, over three decades after its proposal, compatible with the latest observational constraints \cite{Ade:2015lrj}.

The aim of this work is to study possible generalisations of the underlying $f(R)$ theory in $D>4$ dimensions; it is based on \cite{SantiagoThesis}. Recently, there has been further research in this direction \cite{Ketov:2017aau}, which shares some of the conclusions of the present work,  without addressing the important aspect of moduli stabilisation. Whenever higher-dimensional theories are compactified, deformation modes of the internal manifold enter the four-dimensional effective field theory (EFT) as additional scalar fields. Usually those fields must be stabilised in a suitable way to not cause a variety of problems. We study the interplay between inflation from $f(R)$ gravity in higher dimensional spacetimes and moduli stabilisation using a simple toy model. 

We show that, without ingredients other than the gravitational action and a cosmological constant, the potential is generically unstable along the direction of the volume of the compact space. Following the original idea of Freund and Rubin \cite{Freund:1980xh}, we demonstrate that non-vanishing two-form flux on the compact space can lead to sufficiently stable minima with a Minkowski or de Sitter space-time in four dimensions. However, we show that there are no stable inflationary trajectories ending in those minima. While for large values of the scalar field $\phi$ the potential features a plateau -- as in the original Starobinsky model -- this plateau is always unstable in the direction of the volume modulus. Finally, we propose a solution to this problem using a more general $p$-form flux background on the compact space. This allows us to separate moduli stabilisation from the inflationary dynamics.

This work fits well in a line with previous studies of plateau inflation in higher-dimensional theories. For example, the authors of \cite{Burgess:2016ygs} use a similarly simple toy model of moduli stabilisation to investigate its interplay with inflationary theories. Moreover, the past decade has seen substantial progress in string theory implementations of plateau inflation models, also including the study of moduli stability, cf.~\cite{Conlon:2005jm,Cicoli:2008gp,Cicoli:2011ct,Broy:2015zba,Burgess:2016owb,Cicoli:2016chb,Broy:2014xwa}.

The remainder of this paper is organised as follows. In Section 2 we first give the ansatz for the $D$-dimensional $f(R)$ theory. Second, we give the resulting four-dimensional action for the two involved scalar fields in the Einstein frame, after compactification on a sphere. Third, we demonstrate that two-form flux cannot sufficiently stabilise the volume modulus during inflation. Finally, we solve this problem by introducing $p$-form flux on the sphere and discuss the ensuing observational footprint of the model. In Section 3 we conclude, and compile the details regarding the main result, which is the four-dimensional action in Einstein frame, in Appendix A.

%

\section{Starobinsky's model in $D$ dimensions}

The starting point of our discussion is a generalisation of Starobinsky's model in $D$ space-time dimensions. Following \cite{SantiagoThesis} the $D$-dimensional action features an Einstein-Hilbert term, a higher-order curvature term, a cosmological constant, and the kinetic term of a $(p-1)$-form gauge potential. In total, we have 
\begin{align}
S = \frac{M^{D-2}}{2} \int \text d^{D}X \sqrt{-g}  \left(R + \alpha  R^{n}  - 2 M^{2} \Lambda -|F_p|^2 \right)\,,
\label{theactionDn1}
\end{align}
where
\begin{align}
|F_p|^2= g^{M_1N_1} \cdots g^{M_p N_p}F_{M_1\dots M_p}F_{N_1\dots N_p}\,,
\end{align}
and $n$ and $\Lambda$ are treated as free parameters. Moreover, $M$ denotes the $D$-dimensional Planck mass. In the following we are interested in the four-dimensional effective field theory (EFT) after compactification of $D-4$ dimensions on a sphere.\footnote{We choose a sphere because it is a simple example manifold with positive Euler number and a single volume modulus.} $F_p$ only has non-vanishing components in the compact space to satisfy Lorentz invariance in the EFT, so for $D = 4$, $n = 2$, and $\Lambda = 0$ \eqref{theactionDn1} reduces to the standard Starobinsky action.


\subsection{Einstein frame and compactification}
\label{sec:comp}

The action above is written in a $D$-dimensional Jordan frame. To extract the physical predictions of the EFT after compactification of $D-4$ dimensions, an Einstein frame description is particularly useful. The strategy to obtain the desired four-dimensional action is as follows. First, by introducing an auxiliary scalar field $A$ we can remove the term proportional to $R^n$ in \eqref{theactionDn1}. Second, using a conformal transformation of the $D$-dimensional metric, we can transform the result to the $D$-dimensional Einstein frame. Subsequently we compactify the $D-4$ internal dimensions on a sphere. Finally, as the result is again given in a four-dimensional Jordan frame, we perform another conformal transformation to obtain the four-dimensional Einstein frame action of the EFT. The details of this procedure can be found in Appendix \ref{App:AppendixA}. Here we merely state the final result,
\begin{align}
S= \frac12 \int \text{d}^{4}x \sqrt{- g}  \Bigg\{ & R - \frac{1}{2}(D-4) (D-2) \partial_{\mu } \text{ln} \sigma \partial^{\mu} \text{ln} \sigma - \frac{D-1}{D-2}  \partial_{\mu} \text{ln} A \partial^{\mu} \text{ln} A \nonumber \\
&+ 2 \sigma^{2-D}  - \sigma^{4-D} A^{\frac{D}{2-D}} \left[ (n-1) \alpha \left( \frac{A - 1}{n \alpha} \right)^{\frac{n}{n-1}} + 2 \Lambda \right] \nonumber \\
& +V_\text{flux} \Bigg\}\,,
\label{actionplanckfinal}
\end{align}
where $g$ and $R$ now denote the respective four-dimensional quantities. This action is given in terms of four-dimensional natural units, i.e., we have set the four-dimensional Planck mass to unity, $M_\text p = 1$. Also, compared to \eqref{actionplanck} in the appendix we have dropped the hats for convenience. The four-dimensional EFT apparently contains Einstein gravity and two dynamical scalar fields. Here $A$ is the would-be inflaton field analogous to the one in Starobinsky's model and $\sigma$ is the radial modulus of the compact sphere. The canonically normalised variables, $\phi$ and $\Sigma$,  can be deduced from \eqref{actionplanckfinal} and are defined by 
\begin{align}
\label{canonicalchange}
\sigma= e^{ \sqrt{ \frac{2 }{ (D-4)(D-2) }}\Sigma}, \qquad A=e^{\sqrt{\frac{D-2}{D-1}} \phi}\,.
\end{align}
The scalar potential $V(\sigma,A)$ features contributions from the $D$-dimensional higher-order curvature term, from the integrated curvature of the compact space, and from compact space fluxes. It reads
\begin{align}
V = \frac12 \sigma^{4-D} A^{\frac{D}{2-D}} \left[  (n-1) \alpha \left( \frac{A-1}{n \alpha} \right)^{\frac{n}{n-1}} + 2 \Lambda \right] - \sigma^{2-D}+V_\text{flux}\,,
\label{eq:potentialplanck}
\end{align}
where $V_\text{flux}$ is the potential generated by the non-vanishing integral over $|F_p|^2$ on the sphere. It is generally a function of both $A$ and $\sigma$; its form is given below.

If \eqref{eq:potentialplanck} is to have a plateau at large values of $A$, as is typical of four-dimensional Starobinsky inflation, the dimensionality of space-time must be related to the power of the Ricci scalar as follows \cite{SantiagoThesis,Ketov:2017aau},
\begin{align}
D = 2n\,.
\label{eq:Dncond}
\end{align}
We stress that the violation of this condition does not exclude the existence of a flat patch of the potential where inflation can take place. However, in the remainder of the paper we consider setups that feature an infinite plateau in the $A$ direction, for which \eqref{eq:Dncond} is a necessary (but not sufficient) condition. As argued below, the stability of this plateau places non-trivial constraints on the functional form of $V_\text{flux}$.

Before we analyse in detail the interplay between the flux stabilisation of the volume modulus and the existence of a stable and flat inflationary trajectory, let us note that, by taking the limit $D \rightarrow 4$ and $n \rightarrow 2$ while setting $V_\text{flux}=\Lambda=0$, one recovers the standard four-dimensional Starobinsky potential
\begin{align}
\lim_{\substack{D \to 4}} V \Bigg|_{\substack{n=2 \\ \Lambda = 0}} =  \frac{1}{8\alpha} \left( 1-A \right)^{2}\,,
\end{align}
in terms of the non-canonical variable $A$.


\subsection{Volume stabilisation with two-form fluxes}
\label{sec:twoform}

One crucial observation following from the result \eqref{eq:potentialplanck} is that, if $V_\text{flux} = 0$, the theory always has a runaway direction towards $\sigma \to 0$. To arrange for a (meta-)stable minimum of the volume modulus, we can turn on fluxes in the compact space which contribute to the four-dimensional scalar potential. Like in the original Freund-Rubin compactification \cite{Freund:1980xh} (see also  \cite{Douglas:2006es,Denef:2007pq} for a relation to string theory), we may try to employ two-form field strengths. In that case the last term of our starting action \eqref{theactionDn1} reads
\begin{align}
S \supset - \frac{M^{D-2}}{2}  \int \text d^{D}X \sqrt{-g}g^{M N} g^{P Q} F_{MP} F_{NQ}\,,
\end{align}
which, upon dimensional reduction, gives rise to
\begin{align}
V_\text{flux}= \frac12 f^{2} \sigma^{-D} A^{-\frac{D-4}{D-2}}
\label{eq:V2flux}
\end{align}
in the four-dimensional Einstein frame. The integer flux constant $f$ is defined in \eqref{eq:definef}. Thus, the full scalar potential in this case, assuming $D=2n$, reads
\begin{align}
\begin{split}
V(\sigma,A) &= \frac12 \sigma^{2(2-n)} \left[ (n-1) \alpha \left( \frac{1 - A^{-1}}{n \alpha} \right)^{\frac{n}{n-1}}+ 2 A^{ - \frac{n}{n-1}} \Lambda \right] \\ 
&\;\;\;\;+ \frac{1}{2} f^{2} \sigma^{-2n} A^{-\frac{n-2}{n-1}} - \sigma^{2(1-n)}\,.
\label{potentialplanck2n}
\end{split}
\end{align}
We may now study whether this potential has a sufficiently stable minimum with vanishing (four-dimensional) cosmological constant and a stable inflationary trajectory.

A vacuum with the desired properties seems to exist in a limited region of parameter space for any value of $n$. For example, with $n=3$ one finds after solving ${\partial_\sigma V = \partial_A V = V = 0}$,
\begin{align}
\sigma_0^4 = \frac{f^4 + f^2 \lambda}{2}\,, \quad A_0 = \frac{f^4 - f^2 \lambda}{24 \alpha}\,, \quad \Lambda = \frac{- f^4 + f^2 \lambda + 96 \alpha}{72 \alpha (f^2+ \lambda)}\,,
\end{align}
with $\lambda = \sqrt{f^4 - 48 \alpha}$. Thus, the existence of a post-inflationary vacuum implies the parameter constraint $f^4 > 48 \alpha$.

Inflation, however, seems challenging to realise. One can check that for any $n$, the only potentially viable inflationary trajectory in the potential \eqref{potentialplanck2n} is along the coordinate $A$ \cite{SantiagoThesis}. We can evaluate the potential for large values of $A$ as follows, 
\begin{align}
V_\text{lim} = \lim_{\substack{A \to \infty}} V = \frac12 \sigma^{2-2n}\left[\sigma^2(n-1)\alpha (n \alpha)^{\frac{n}{1-n}} -2 \right]\,.
\end{align}
The result does not depend on $A$, so the potential develops a plateau as in the original setup of Starobinsky.
However, the plateau is always unstable in the direction of the modulus $\sigma$. In fact, $V_\text{lim}$ has a single local extremum at 
\begin{align}
\sigma_\text{c}^2 = \frac{2 (n \alpha)^{\frac{n}{n-1}}}{\alpha(n-2)}\,,
\end{align}
which does feature a positive value for the scalar potential on the plateau, 
\begin{align}
V_\text{plat} = V_\text{lim}(\sigma_\text c) = \frac{2^{1-n}}{\alpha}  (n-2)^{n-2} n^{-n}\,,
\end{align}
but a mass for $\sigma$ that is always negative for $n>2$,
\begin{align}
\partial_\sigma^2 V_\text{lim}(\sigma_\text c) = -2^{2-n} \alpha^n (n-1) (n-2)^n (n \alpha)^\frac{n^2}{1-n}\,.
\end{align}
This leads us to exclude the possibility of Starobinsky inflation in $D > 4$ dimensions in cases where the radial modulus of the compact dimensions is stabilised by two-form flux. This is ultimately due to the fact that, as a result of the dimensional reduction and conformal transformation, the flux term in \eqref{potentialplanck2n} depends inversely on $A$. Hence, for large $A$ the crucial stabilising term is eliminated. In the following, we show how this problem can be avoided in a more general flux background.

\subsection{Volume stabilisation with $p$-form fluxes}
\label{sec:pform}

In order to disentangle problem of moduli stabilisation from the potential of the would-be inflaton field, one can consider the more general case of stabilisation via $p$-form fluxes with $p>2$. The corresponding term in the original action is then
\begin{align}
S\supset  - \frac{M^{D-2}}{2} \int \text d^{D}X \sqrt{-g} g^{M_1N_1}...  g^{M_p N_p}F_{M_1...M_p}F_{N_1...N_p}.
\label{eq:Spform}
\end{align}
After noticing that the source of difficulties in the two-form case is the $A$ dependence in \eqref{eq:V2flux}, we consider flux terms that are invariant under the Legendre-Weyl transformation that recasts the $D$-dimensional action into the Einstein frame.\footnote{For details cf.~Appendix \ref{App:AppendixA}.} This implies a link between the rank of the $p$-form and the dimensionality of space-time,
\begin{align}
D=2p\,.
\end{align}
This degree of flux is only possible if $D\ge8$, since it must be $p \in \mathbb N$ and $p \le D-4$.

As shown in Appendix \ref{App:AppendixA}, upon dimensional reduction \eqref{eq:Spform} gives rise to the following term in the four-dimensional Einstein-frame action,
\begin{align}
V_\text{flux} = \frac12 f^2 \sigma^{-2D+4}=\frac12 f^2 \sigma^{4-4n}\,,
\end{align}
where the last equality follows from imposing $D=2n$. As advertised, this stabilising term is independent of $A$. The full scalar potential then reads
\begin{align}
V(\sigma,A)  = \frac12 \sigma^{2(2-n)} \left[ (n-1) \alpha \left( \frac{1 - A^{-1}}{n \alpha} \right)^{\frac{n}{n-1}}+ 2 A^{ - \frac{n}{n-1}} \Lambda \right]  + \frac12 f^2 \sigma^{4-4n} -  \sigma^{2(1-n)}.
\label{eq:potpform}
\end{align}
Again we find a stable Minkowski vacuum for any $n$, given by 
\begin{align}
\sigma_0 = \left(\frac{f^2 n}{2}\right)^\frac{1}{2n-2}\, , \quad A_0 =\frac{f^2}{f^2 - 2^n \alpha} \,, \quad \Lambda = \frac{n-1}{n}\left(\frac{f^2 n}{2}-2^{n-1}\alpha n \right)^\frac{1}{1-n}\,.
\label{eq:minp}
\end{align}

As in Section \ref{sec:twoform} we may look for the possibility of a plateau at large values of $A$. Indeed one finds in the $A \to \infty$ limit
\begin{align}
V = \frac12 f^2 \sigma^{4-4n}-\sigma^{2-2n} +\frac{(n-1)(n \alpha)^\frac{1}{1-n}}{2 n}\sigma^{4-2n}+\mathcal{O}(A^{-1})\,.
\end{align}
In this regime the volume modulus actually develops a local minimum at $\sigma_\text c$, defined by
\begin{align}
f^2=\sigma_\text c^{2n-2}\left(1+\frac{2-n}{2n}(n \alpha)^\frac{1}{1-n} \sigma_\text c^2\right)\,,
\label{eq:sigmac}
\end{align}
which implies that the height of the plateau at large $A$ is given by
\begin{align}
V_\text{plat} = \frac{1}{4}\sigma_\text c^{2-2n}\left(-2+(n \alpha)^\frac{1}{1-n} \sigma_\text c^2 \right)\,.
\end{align}
This situation is different from the one with two-from fluxes in Section \ref{sec:twoform}. The plateau is actually stable in a certain parameter regime, since the mass of $\sigma$ can be positive and large compared to the inflationary energy scale. In particular, one finds for the mass of the canonically normalised modulus at $\sigma_\text c$,
\begin{align}
m^2_\Sigma=\sigma_\text c^{2-2n}\left( \frac{2n-2}{n-2}-(n \alpha)^\frac{1}{1-n}\sigma_\text c^2\right)\,.
\end{align} 
Requiring the inflationary dynamics to be described by a single-field system, i.e., imposing that $\sigma$ can be integrated out consistently, leads to the following two constraints on the parameters of the model,
\begin{align}
m^2_\Sigma>0\,, \quad \frac{m^2_\Sigma}{V_\text{plat}} \gg 1\,.
\end{align}
The latter constraint comes from the requirement that the dynamics of $\sigma$ are negligible during the inflationary epoch. 
These constraints imply a tuning of the parameters such that
\begin{align}
2 <  \sigma_\text c^2 (n\alpha)^\frac{1}{1-n} < \frac{2n-12/5}{n-2}\,.
\label{eq:tuning}
\end{align}
With $n>3$, as has to be the case in our setup, one finds $\frac{2n-12/5}{n-2}<4$. Moreover, note that \eqref{eq:minp}, \eqref{eq:sigmac}, and \eqref{eq:tuning} imply that in the desired parameter regime $\sigma_0 \approx \sigma_\text c$. This means that the back-reaction of the inflationary energy density on the expectation value of the volume modulus is negligible.

\paragraph{Validity of the four-dimensional EFT\\}
In order to evaluate the validity of the four-dimensional EFT one must compare the energy scales in the problem to the Kaluza-Klein (KK) scale of the compactification. Let us expand
\begin{equation}
\sigma_\text c^2 (n\alpha)^\frac{1}{1-n} \equiv 2+\delta\,,
\end{equation}
where $\delta\ll1$. One can then show that 
\begin{equation}
V_\text{plat}=\frac{1}{4}\sigma_\text{c} ^{2-2n} \delta
\qquad
\text{and}
\qquad
m^2_\Sigma= \sigma_\text{c} ^{2-2n} \frac{2}{n-2}+\mathcal{O}(\delta)\,.
\end{equation}
For the four-dimensional description to be valid both energy scales must be below the KK scale, $V_\text{plat}\ll m^2_\Sigma \ll M_\text{KK}^2$, which is given by
\begin{equation}
M_\text{KK}\simeq\frac{1}{\sigma}\,.
\end{equation}

Since one can tune $\delta\ll1$ it automatically follows that the Hubble parameter during inflation is parametrically smaller than the square of the  KK scale. The situation of $m_\Sigma$ is more subtle since for $D= 2n\ge 8$ one finds
\begin{equation}
\frac{m_\Sigma^2}{M_\text{KK}^2}\simeq\frac{2}{n-2} \lesssim 1\,.
\end{equation}
We therefore conclude that the mass of the volume mode is below, but very close to the KK scale. We note that by tuning the dimensionality of space-time the ratio can be made smaller but that a hierarchical separation is hard to achieve. This renders the moduli stabilisation physics discussed above vulnerable to corrections coming from higher-dimensional physics.

\paragraph{Inflationary footprint\\}

If \eqref{eq:tuning} is fulfilled we can describe inflation in terms of a single-field Lagrangian with the scalar potential $V (A) \approx V(\sigma_0, A)$ to very high accuracy. We can then determine the observational footprint of the model as follows. The inflationary potential in terms of the canonical variable $\phi$, defined in \eqref{canonicalchange}, reads
\begin{align}
V_\text{inf}= \mathcal{C}_1+\mathcal{C}_2  e^{-\frac{n}{n-1}\kappa \phi}+\mathcal{C}_3 \left(1-e^{-\kappa \phi}\right)^\frac{n}{n-1}\, ,
\end{align}
where $\kappa \equiv \sqrt{\frac{2n-2}{2n-1}}$ and the $\mathcal{C}_i$ can be read off from \eqref{eq:potpform} after setting $\sigma = \sigma_0$. Notice that the value of $\kappa$ plays a pivotal role in the determination of the observables for this class of potentials. For interesting cases one finds
\begin{align}
\kappa|_{D=8}=\sqrt{\frac{6}{7}}\,,\quad \kappa|_{D=10} =\frac{2\sqrt{2}}{3}\,.
\end{align}

\begin{figure}[t!]
	\centering
	\includegraphics[width=0.8\textwidth]{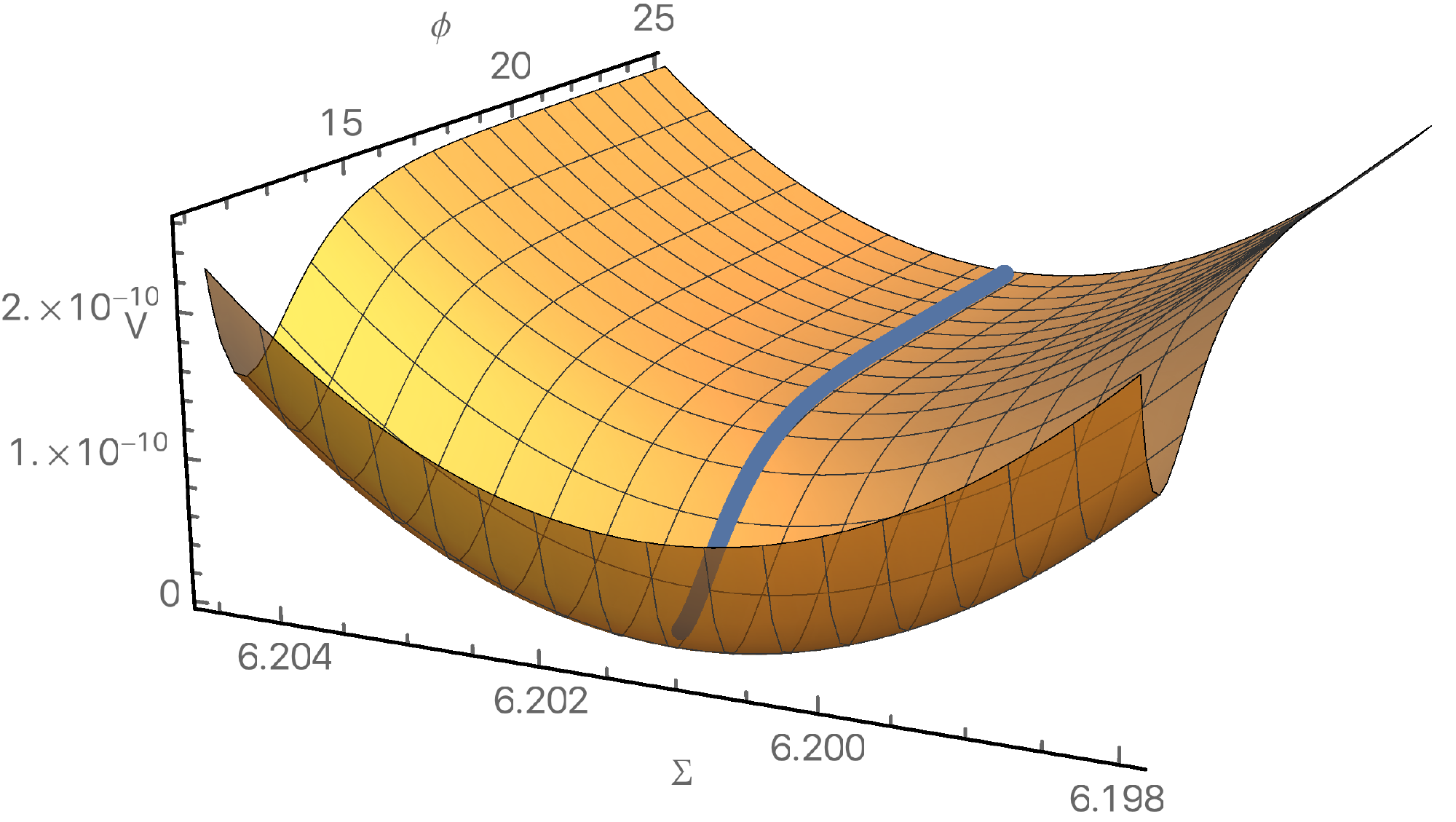}
 \caption{Inflationary potential in terms of the canonically normalised fields $\Sigma$ and $\phi$. We have chosen $D = 10$, leading to $n = p = 5$, as well as $f = 100$ and $\alpha = 312.496$. The blue line indicates a possible slow-roll trajectory starting on the plateau. Notice the different scales on the $\phi$ and $\Sigma$ axes, indicating the large mass difference.}
	\label{fig:D10examples}
\end{figure}

As mentioned above, the single-field regime of this setup can be reached whenever the conditions \eqref{eq:tuning} are imposed. The closer the parameter choice is to saturating the lower bound on the left-hand side of \eqref{eq:tuning}, the more robust the mass hierarchy between volume modulus and the inflaton becomes. Furthermore, the correct normalisation of the scalar perturbations requires that at horizon exit $V \sim 10^{-10}$ in Planck units. Hence the closer the parameter choice is to saturating the lower bound, the smaller the radius of the compact space $\sigma_0$, and consequently the smaller the required values of the parameters $f$ and $\alpha$. For illustration, Figure \ref{fig:D10examples} depicts one correctly normalised example with a large mass hierarchy.

In what concerns CMB observables, even in $D > 4$ dimensions, one recovers values similar to the well-known ones for Starobinsky-type potentials, namely
\begin{align}
n_\text s\approx 1-\frac{2}{N_e}\,,\quad r\approx\frac{9}{\kappa^2 N_e^2}\,,
\end{align}
where $N_e$ denotes the number of $e$-folds of expansion. Therefore, while these models lie at the centre of the Planck 1$-\sigma$ region \cite{Ade:2015lrj}, they are essentially indistinguishable among themselves and also from the four-dimensional Starobinsky model with $\kappa=\sqrt{2/3}$.

%

\section{Discussion}

In this paper we have explored the relation between an $R+\alpha R^n$ gravitational theory in a $D$-dimensional space-time and the occurrence of inflation in four dimensions. This work constitutes an obvious extension of the Starobinsky model of inflation. 

Using the example of a sphere with a single volume modulus, we have found that the stabilisation and dynamics of the extra-dimensional manifold is closely connected to inflation and that disentangling the two requires judicious choices of the model parameters. This situation is analogous to well known results in string inflation, where the interplay between inflation and moduli stabilisation has been extensively studied over the last decade.

The stand-out feature of the original four-dimensional Starobinsky proposal, apart from the fact that after 30 years it is nowhere near being excluded by CMB data, is the existence of an infinite plateau at large field values. In our $D$-dimensional case, demanding the scalar potential in the Einstein frame to have a similar plateau constrains the form of the initial gravitational action to $f(R)=R+\alpha R^{D/2}$. Requiring stability of the compact space during inflation further constrains the form of the action, determining the extra degrees of freedom that can be present in the UV limit.  More concretely, it excludes stabilisation of the compact space with a two-form field strength, as in Freund-Rubin compactifications. Instead, one may stabilise the volume via $p$-forms, where the rank $p$ is related to the dimensionality of space-time, $p=D/2$. This last constraint combined with four-dimensional Lorentz invariance forces us to consider spaces of even dimensionality with $D\ge8$. Once all these conditions can be met, it is possible to tune the microscopic parameters -- such as the amount of flux, the $D$-dimensional cosmological constant, and the strength of the $R^n$ term -- to generate viable models of single-field inflation, compatible with the latest observational constraints, that exit into a viable post-inflationary minimum.

%

\section*{Acknowledgments}
This work is partially supported by the grants FPA2012-32828 from the MINECO, the ERC Advanced Grant SPLE under contract ERC-2012-ADG-20120216-320421 and the grant SEV-2012-0249 of the ``Centro de Excelencia Severo Ochoa" Programme.

%

\appendix
\section{Appendix}


\subsection{Derivation of the four-dimensional Einstein frame action}
\label{App:AppendixA}

Here we perform the series of transformations that take the action from the $D$-dimensional $f(R)$ frame of  \eqref{theactionDn1} to its four-dimensional Einstein frame form, cf.~$\eqref{actionplanckfinal}$. As usual, the Einstein frame is defined as the frame in which the gravitational action takes the Einstein-Hilbert form, the Jordan frame is the one in which the Ricci scalar appears multiplied by a function of a scalar field and the $f(R)$ frame is the one in which the gravitational part of the action is expressed as a (non-linear) function of the Ricci scalar. In this paper, in a slight abuse of nomenclature, we refer to the Jordan and $f(R)$ frames indiscriminately. 

Let us first focus on the pure gravity part of the action in order to write it in the Einstein-Hilbert form in $D$ dimensions.  We decompose \eqref{theactionDn1} into $S= S_\text{grav} + S_\text{matt}$, where
\begin{align}
S_\text{grav}  = \frac{M^{D-2}}{2} \int \text{d}^{D}X \sqrt{-g}  \left( R + \alpha  R^{n} \right)\,,
\label{eq:SG}
\end{align}
and
\begin{align}
S_\text{matt}  = \frac{M^{D-2}}{2} \int \text{d}^{D}X \sqrt{-g} \left( - 2 M^2 \Lambda - g^{M_1 N_1}...g^{M_p N_p}F_{M_1...M_p}F_{N_1...N_p} \right)\,.
\label{eq:Smatter}
\end{align}
The $D$-dimensional cosmological constant $\Lambda$ is dimensionless and the field strength $p$-forms have mass dimension one.

Let us introduce an auxiliary field $\chi$ with mass dimension $2$, and write the action as \cite{DeFelice:2010aj}
\begin{align}
S =  \frac{M^{D-2}}{2} \int \text{d}^{D}X \sqrt{-g} \left[ f(\chi) + \frac{ \partial f(\chi)}{\partial \chi } \left( R - \chi \right) \right]\,.
\label{auxiliaraction}
\end{align}
Note that, at this level, $\chi$ is a genuine auxiliary field because its action has no time derivatives. The equation of motion for $\chi$ following from this action is
\begin{align}
\frac{ \partial^{2} f(\chi)}{\partial \chi^{2}} \left( R - \chi \right) = 0 \quad \Rightarrow \quad R = \chi\,,
\label{eomchi}
\end{align}
where we used that $\frac{ \partial^{2} f(\chi)}{\partial \chi^{2} } = n (n-1) \alpha \chi^{n-2} \neq 0$. This implies that the action $\eqref{auxiliaraction}$ is trivially equivalent to $\eqref{eq:SG}$. After defining the dimensionless field $A$ via
\begin{align}
A \equiv \frac{ \partial f(\chi)}{\partial \chi} = 1 + n \alpha \chi^{n-1}\,,
\label{Afield}
\end{align}
we can express the action in $\eqref{auxiliaraction}$ as
\begin{align}
S = \frac{M^{D-2}}{2} \int \text{d}^{D}X \sqrt{-g} \left( A R - Z(A) \right)\,,
\label{auxiliaraction2}
\end{align}
where
\begin{align}
\label{Z(A)}
Z(A) \equiv \chi (A) A - f(\chi(A)) = (n-1) \alpha \chi^{n}(A)=(n-1) \alpha \left( \frac{A - 1}{n \alpha} \right)^{\frac{n}{n-1}}\,.
\end{align}
It is useful to write all dimensionful parameters in terms of the $D$-dimensional Planck mass $M$, so we rescale
\begin{align}
\alpha  \to  M^{2-2n} \alpha \,, \quad  Z(A) \to M^2 Z(A)\,.
\label{alpha}
\end{align}
From this point onwards $\alpha$ (like $\Lambda$) is dimensionless. So far, the total action is thus 
\begin{align}
S =  \frac{M^{D-2}}{2} \int \text{d}^{D}X \sqrt{-g} \left( A R - M^2 Z(A) - 2 M^2 \Lambda - g^{M_1 N_1}\dots g^{M_n N_n}F_{M_1 \dots M_n}F_{N_1\dots N_n} \right)\,.
\label{theactionDn}
\end{align}

In order to transform this to the $D$-dimensional Einstein frame, we perform a conformal transformation and write the action in terms of the metric $\tilde{g}$ defined by
\begin{align}
\label{conformal1}
g_{M N}= \Omega^{-2} \tilde{g}_{MN}\,, \quad g^{M N}= \Omega^{2} \tilde{g}^{MN}\,, \quad\sqrt{-g_{(D)}} =\Omega^{-D}\sqrt{-\tilde{g}_{(D)}}\, .
\end{align}
One can show that under a Weyl rescaling $R$ transforms as \cite{Fujii:2003pa}
\begin{align}
R = \Omega^{2}\left[ \tilde{R}+2(D-1)\tilde{g}^{M N}\tilde{\nabla}_{M}\tilde{\nabla}_{N} \text{ln} \Omega - (D-1)(D-2)\tilde{g}^{M N}(\partial_{M} \text{ln} \Omega)(\partial_{N} \text{ln} \Omega)  \right]\, .
\label{conformalR1}
\end{align}
where quantities with a tilde are understood with respect to $\tilde{g}_{MN}$. In order for $\tilde{g}$ to be the $D$-dimensional Einstein frame metric it must be
\begin{align}
\Omega=A^{\frac{1}{D-2}}\,.
\label{fix1}
\end{align}
Then the action in the $D$-dimensional Einstein frame reads
\begin{align}
\label{actioneinsteinframe1}
\nonumber S= \frac{M^{D-2}}{2} \int \text{d}^{D}X \sqrt{-\tilde{g}} \biggl[ & \tilde{R}+2 \frac{D-1}{D-2}  \tilde{g}^{M N} \tilde{\nabla}_{M}\tilde{\nabla}_{N} \ln A - M^2 A^{\frac{D}{2-D}} Z(A)  \\ 
\nonumber & - \frac{D-1}{D-2} \tilde{g}^{M N} \partial_{M} \ln A \partial_{N} \ln A - 2 M^2 A^{\frac{D}{2-D}} \Lambda\\ 
& - A^{\frac{2 p-D}{D-2}}  \tilde{g}^{M_1 N_1}\dots \tilde{g}^{M_p\dots N_p}F_{M_1\dots M_p}F_{N_1\dots N_p} \biggr]\,.
\end{align}
In what follows we ignore the total derivative $\tilde{\nabla}^2 \ln A$.

Space-time in our framework is described by a $D$-dimensional manifold $\mathcal{M}_{D}$ equipped with the metric $\tilde{g}_{M N}$. This $D$-dimensional manifold can be factorised into a four-dimensional manifold $\mathcal{M}_{4}$ and a $(D-4)$-dimensional manifold $\mathcal{M}_{D-4}$,
\begin{align}
\mathcal{M}_{D} = \mathcal{M}_{4} \times \mathcal{M}_{D-4}\,,
\label{dmanifold}
\end{align}
such that the metric $g_{M N}$ can be written in a block-diagonal form,
\begin{align}
\label{metricds}
\text{d} s^{2} = \tilde{g}_{M N} \text{d} X^{M} \text{d} X^{N} = g_{\mu \nu} \text{d} x^{\mu} \text{d} x^{\nu} + \sigma^{2} g_{m n} \text{d} y^{m} \text{d} y^{n},
\end{align}
with $M,N = \{0,\dots, D-1\}$; $ \mu,\nu =\{ 0,\dots, 3\} $, and $ m,n = \{4,\dots, D-1\}$. With this notation we choose the metric $g_{mn}$ to have unit volume such that the physical volume of the compact space is determined by the dimensionless scalar field $\sigma$. Moreover, we choose the compact space to be a sphere. In what follows, we assume that $\sigma=\sigma(x)$ and $A=A(x)$, i.e., the scalars have constant profiles in the compact space. We further assume, in line with Freund-Rubin stabilisation, that the $p$-form fluxes are non-vanishing in the compact space but vanishing in the external dimensions, thereby preserving four-dimensional Lorentz invariance. This translates into
\begin{align}
\begin{split}
\mathcal{L}\supset &-\frac{M^{D-2}}{2} A^{\frac{2 p-D}{D-2}}  \tilde{g}^{M_1 N_1}\dots \tilde{g}^{M_p\dots N_p}F_{M_1\dots M_p}F_{N_1 \dots N_p}  \\ 
= &-\frac{M^{D-2}}{2} A^{\frac{2 p-D}{D-2}}  \sigma^{-2 p} g^{m_1 n_1}\dots g^{m_p\dots n_p}F_{m_1\dots m_p}F_{n_1\dots n_p}\, .
\end{split}
\end{align}
Given the block-diagonal form in \eqref{metricds} one may factorise the determinant,
\begin{align}
\label{determinant}
\sqrt{-\tilde{g}_{(D)}}=\sigma^{D-4}  \sqrt{-g_{(4)}} \sqrt{g_{(D-4)}},
\end{align}
where $g_{(4)}$ and  $g_{(D-4)}$ are the determinants of $g_{\mu \nu}$ and $g_{mn}$, respectively. Then the $D$-dimensional Ricci scalar decomposes as follows,
\begin{align}
 \tilde{R}_{(D)}  =  R_{(4)} + \frac{R_{(D-4)}}{\sigma^2} - 2(D-4)  g^{\mu \nu} \frac{\nabla_{\mu } \nabla_{\nu}\sigma}{\sigma} - \left[ (D-4)^{2}-(D-4)\right]g^{\mu \nu}\frac{ \partial_{\mu }\sigma \partial_{\nu} \sigma}{\sigma^2}\,,
\label{ricciD}
\end{align}
where $R_{(4)}$ is the curvature of the four-dimensional metric $g_{\mu\nu}$, $R_{(D-4)}$ is the curvature of the $(D-4)$-dimensional metric $g_{m n}$ and $\nabla_{\mu}$ is the four-dimensional covariant derivative with respect to $g_{\mu\nu}$. Under this decomposition the $D$-dimensional Einstein-Hilbert action becomes
\begin{align}
\label{einsteinhilbertdecomposed}
\nonumber S = & \;\; \frac{M^{D-2}}{2}  \int \text{d}^4 x \text{d}^{D-4} y \sqrt{-g_{(4)}}  \sqrt{g_{(D-4)}} \sigma^{D-4}\biggl\{ R_{(4)} + \frac{R_{(D-4)}}{\sigma^2} \\
& - 2(D-4) g^{\mu \nu} \frac{ \nabla_{\mu } \nabla_{\nu}\sigma}{\sigma} - \left[ (D-4)^{2}-(D-4)\right]g^{\mu \nu} \frac{\partial_{\mu }\sigma \partial_\nu \sigma}{\sigma ^2} \biggr\}\,.
\end{align}
One can, at this point, perform the integral over the $(D-4)$-dimensional internal space. We remember that 
\begin{align}
\label{genus1}
\int \text d^{D-4} y \sqrt{g_{(D-4)}}  R_{(D-4)}= \chi M^{6-D}= 2 M^{6-D}\,, 
\end{align}
for the Euler characteristic of the compact sphere. Moreover, the volume of the compact space is given by
\begin{align}
\label{volume}
\mathcal{V}= \sigma^{D-4} \int \text d^{D-4}y  \sqrt{g_{(D-4)}}  = M^{4-D} \sigma^{D-4}\,,
\end{align}
where we used that
\begin{align}
\label{integralcompact}
\int \text d^{D-4}y \sqrt{g_{(D-4)}}  =  M^{4-D}\,.
\end{align}
Using $\eqref{determinant}$, $\eqref{genus1}$, and $\eqref{volume}$ we can now explicitly perform the integration. For later convenience we multiply and divide by the vacuum expectation value of the field $\sigma_{0}^{D-4}$. This step is necessary to find the relation between the $D$-dimensional Planck mass and the four-dimensional Planck mass. The action in the four-dimensional Jordan frame then becomes
\begin{align}
\label{actioncompactified}
\nonumber S= \frac{M^2 \sigma_0^{D-4}}{2} \int \text{d}^{4}x \sqrt{-g_{(4)}}  \left(\frac{\sigma}{\sigma_{0}}\right)^{D-4}  \biggl\{ &R_{(4)} + M^2 \sigma^{-2} + (D-4)(D-5)  \sigma^{-2} g^{\mu \nu} \partial_{\mu }\sigma \partial_{\nu}\sigma \\  \nonumber
&-\frac{D-1}{D-2} g^{\mu \nu} \partial_{\mu} \ln A \partial_\nu \ln A -  M^2 A^{\frac{D}{2-D}} \left[ Z(A)+ 2 \Lambda \right]  \nonumber \\
 & - M^2 \sigma^{-2p}  A^{\frac{2p-D}{D-2}} f^{2}  \biggr\}.
\end{align}
We have used partial integration on the $\nabla^2 \sigma$ term of in \eqref{einsteinhilbertdecomposed} and defined the dimensionless flux constant $f$ via
\begin{align}\label{eq:definef}
\int \text d^{D-4}y \sqrt{g_{(D-4)}} g^{m_1n_1}\dots g^{m_p n_p} F_{m_1\dots m_p}F_{n_1\dots n_p}\equiv M^{2-D} f^2.
\end{align}

A further conformal transformation is necessary to yield the four-dimensional Einstein frame, so we define the new metric $\hat{g}$ via
\begin{align}
\label{conformal2}
g_{\mu \nu}  =\Omega^{-2}\hat{g}_{\mu \nu}\,, \quad g^{\mu \nu}= \Omega^{2}\hat{g}^{\mu \nu}\,, \quad \sqrt{-g_{(4)}}= \Omega^{-4}\sqrt{-\hat{g}_{(4)}}\,.
\end{align}
$R$ transforms under this conformal transformation as follows,
\begin{align}
\label{conformalR2}
R = \Omega^{2}\left( \hat{R}+6\hat{g}^{\mu \nu}\hat{\nabla}_{\mu}\hat{\nabla}_{\nu}\ln \Omega - 6\hat{g}^{\mu \nu}\partial_{\mu} \ln \Omega \partial_{\nu} \ln \Omega  \right) \,.
\end{align}
Imposing that $\hat{g}$ is the four-dimensional Einstein frame metric fixes
\begin{align}
\label{fix2}
\Omega = \left(\frac{\sigma}{\sigma_{0}}\right)^{\frac{D-4}{2}}\,,
\end{align}
which implies that the action takes the following form,
\begin{align}
\label{actioncompactified}
\nonumber S= \frac{M^2}{2}\sigma_{0}^{D-4} \int \text{d}^{4}x \sqrt{-\hat{g}_{(4)}} \biggl\{ & \hat{R}_{(4)} + M^2 \sigma_0^{D-4} \sigma^{2-D}+ 3(D-4)\hat{g}^{\mu \nu}\hat{\nabla}_{\mu} \hat{\nabla}_{\nu} \text{ln} \sigma \\ 
\nonumber & - \frac{3}{2}(D-4)^2 \hat{g}^{\mu \nu}\partial_{\mu} \ln \sigma \partial_{\nu} \ln \sigma + (D-4)(D-5) \hat{g}^{\mu \nu} \partial_{\mu } \ln \sigma \partial_{\nu} \ln \sigma \\ 
\nonumber &-\frac{D-1}{D-2}  \hat{g}^{\mu \nu} \partial_{\mu} \ln A \partial_{\nu} \ln A  -M^2 \frac{\sigma^{4-D}}{\sigma_{0}^{4-D}} A^{\frac{D}{2-D}} \left[ Z(A)+ 2 \Lambda \right] \\
& - M^2 \sigma^{-D} A^{-\frac{D-4}{D-2}} \sigma_{0}^{D-4} f^{2} \biggr\}\,,
\end{align}
The last term in the first line is a total derivative and can be neglected. Defining the four-dimensional Planck mass in terms of $M$ and $\mathcal V$ as follows,
\begin{align}
M_\text p^2\equiv M^2 \sigma_0^{D-4}=M^2 \mathcal{V}_0\,,
\end{align}
allows us to write the action in its most useful form,
\begin{align}
\label{actionplanck}
S= \frac{M_\text p^2}{2} \int \text{d}^{4}x \sqrt{-\hat{g}_{(4)}}   \Biggl\{ & \hat{R}_{(4)} - \frac{1}{2}(D-4) (D-2) \hat{g}^{\mu \nu} \partial_{\mu } \ln \sigma \partial_{\nu} \ln \sigma \nonumber \\ \nonumber
&-\frac{D-1}{D-2}  \hat{g}^{\mu \nu} \partial_{\mu} \ln A \partial_{\nu} \ln A + M_\text p^{2}\sigma^{2-D} \\ \nonumber
& -M_\text p^2 \sigma^{4-D} A^{\frac{D}{2-D}}  \left[ (n-1) \alpha \left( \frac{A - 1}{n \alpha} \right)^{\frac{n}{n-1}}+ 2 \Lambda \right] \\ 
& - M_\text p^2 \sigma^{4-2p-D} A^{\frac{2p-D}{D-2}}  f^{2} \Biggr\},
\end{align}
where have used \eqref{Z(A)}. This is the result given in Section \ref{sec:comp}, where we set $M_\text p = 1$ and omit the hats and indices on $g$ and $R$ for clarity.

\end{document}